\documentclass[twocolumn,showpacs,preprintnumbers,amsmath,amssymb,superscriptaddress,nofootinbib,english]{revtex4-1}
\pdfoutput=1
\usepackage{times,amsmath,amsfonts,amssymb,epstopdf}
\usepackage{graphicx}
\usepackage{dcolumn}
\usepackage{bm}
\usepackage{epsfig}
\usepackage{graphicx}
\usepackage{hyperref}
\usepackage[usenames]{color}
\usepackage{url}
\usepackage[normalem]{ulem}
\usepackage[T1]{fontenc}
\usepackage[dvipsnames]{xcolor}
\usepackage[titletoc]{appendix}

\def\be{\begin{equation}}
\def\ee{\end{equation}}
\def\ba{\begin{eqnarray}}
\def\ea{\end{eqnarray}}
\begin{document}

\title{Speeding up $N$-body simulations of modified gravity: Chameleon screening models}

\author{Sownak~Bose}
\email[Email address: ]{sownak.bose@durham.ac.uk}
\affiliation{Institute for Computational Cosmology, Department of Physics, Durham University, Durham DH1 3LE, UK}

\author{Baojiu~Li}
\email[Email address: ]{baojiu.li@durham.ac.uk}
\affiliation{Institute for Computational Cosmology, Department of Physics, Durham University, Durham DH1 3LE, UK}

\author{Alexandre~Barreira}
\affiliation{Max-Planck-Institut f{\"u}r Astrophysik, Karl-Schwarzschild-Str. 1, 85741 Garching, Germany}

\author{Jian-hua~He}
\affiliation{Institute for Computational Cosmology, Department of Physics, Durham University, Durham DH1 3LE, UK}

\author{Wojciech~A.~Hellwing}
\affiliation{Institute of Cosmology and Gravitation, University of Portsmouth, Portsmouth PO1 3FX, UK}
\affiliation{Janusz Gil Institute of Astronomy, University of Zielona G\'ora, ul.Szafrana 2, 65-516 Zielona G\'ora, Poland}

\author{Kazuya~Koyama}
\affiliation{Institute of Cosmology and Gravitation, University of Portsmouth, Portsmouth PO1 3FX, UK}

\author{Claudio~Llinares}
\affiliation{Institute for Computational Cosmology, Department of Physics, Durham University, Durham DH1 3LE, UK}

\author{Gong-Bo Zhao}
\affiliation{National Astronomy Observatories, Chinese Academy of Science, Beijing, 100012, P.R. China}
\affiliation{Institute of Cosmology and Gravitation, University of Portsmouth, Portsmouth PO1 3FX, UK}

\date{\today}

\begin{abstract}

We describe and demonstrate the potential of a new and very efficient
method for simulating certain classes of modified gravity theories,
such as the widely studied $f(R)$ gravity models. High resolution
simulations for such models are currently very slow due to the highly
nonlinear partial differential equation that needs to be solved
exactly to predict the modified gravitational force. This nonlinearity
is partly inherent, but is also exacerbated by the specific numerical
algorithm used, which employs a variable redefinition to prevent
numerical instabilities. The standard Newton-Gauss-Seidel iterative
method used to tackle this problem has a poor convergence rate. Our
new method not only avoids this, but also allows the discretised
equation to be written in a form that is analytically solvable. We
show that this new method greatly improves the performance and
efficiency of $f(R)$ simulations. For example, a test simulation with
$512^3$ particles in a box of size $512 \, \rm{Mpc}/h$ is now 5 times
faster than before, while a Millennium-resolution simulation for
$f(R)$ gravity is estimated to be more than 20 times faster than with
the old method. Our new implementation will be particularly useful for
running very high resolution, large-sized simulations which, to date,
are only possible for the standard model, and also makes it feasible
to run large numbers of lower resolution simulations for covariance
analyses. We hope that the method will bring us to a new era for
precision cosmological tests of gravity.

\end{abstract}

\pacs{}

\maketitle

\section{Introduction}

\label{sect:introduction}

Modified gravity theories \cite{cfps2011,jls2016} are popular
alternatives to the cosmological constant and dark energy models
\cite{cst2006} to explain the observed accelerating expansion of our
Universe
\cite{gsetal2010,pretal2009,bbetal2011,rsetal2012,hletal2012,rmetal2009}. Rather
than invoking a cosmological constant ($\Lambda$), or a new energy
component to drive the dynamics of the cosmos, these theories suggest
that the Universe contains only normal and dark matter (which is often
assumed as cold dark matter, or CDM), but the law of gravitation
deviates from that prescribed by Einstein's General Relativity (GR) on
large scales, resulting in an acceleration of the expansion rate.

Since the law of gravity is universal, deviations from GR on large
scales are often associated with changes in the behaviour on small
scales. Any such small scale changes, however, must be vanishingly
small due to the strong constraints placed by numerous local tests of
gravity \cite{w2014}. Consequently, viable modified gravity theories
usually have some mechanism by which such modifications are
suppressed, recovering GR in dense regions like the Solar System,
where those gravity tests have been carried out and their resulting
constraints apply. These are commonly referred to as `screening
mechanisms' in the literature, and are an inherent (instead of an
add-on) property which comes from the dynamics of the theory. The
screening effect implies that gravity behaves differently in different
environments; this environmental dependence is often reflected in
strong nonlinearities in the field equations, which make both
analytical and numerical studies of such theories challenging.

In most theories that are currently being investigated, the
modification to GR boils down to an extra (so-called fifth) force that
is mediated by a new scalar field, and screening in this context means
suppression of the fifth force. In one class of such theories, this is
achieved by a coupling of the scalar field to matter and a nonlinear
self-interaction potential of the scalar field. With appropriate
choices of the coupling and potential, the dynamics of the scalar
field can ensure that, in high density regions, the fifth force it
mediates decays exponentially fast with distance, or becomes extremely
small in its amplitude. Chameleon theories \cite{kw2004,ms2007}, with
$f(R)$ gravity \cite{sf2010} (see also \cite{lb2007,hs2007,bbds2008})
as a representative example, is an instance of the former case, while
the dilaton \cite{bbds2010} and symmetron \cite{hk2010} models belong
to the latter case. Amongst these models, $f(R)$ gravity is currently
the most well-studied case, and there exist numerous works
investigating in detail its predictions for large-scale structure
formation in the nonlinear regime. This has been made possible by the
continuous development of $N$-body simulation codes
\cite[e.g.,][]{o2008,olh2008,sloh2009,lz2009,zmlhf2010,lz2010,zlk2011,zlk2011b,lh2011,lzk2012,lkzl2012,lzlk2012,jblzk2012,lhkzjb2012}.
An efficient code amongst these is {\sc ecosmog} \cite{ecosmog}, based
on the publicly available $N$-body and hydro code {\sc ramses}
\cite{ramses}, which makes large simulations for $f(R)$ gravity
feasible. Using the generic parametrisation for modified gravity
theories \cite{bdl2011,bdlw2012}, {\sc ecosmog} was extended to
incorporate chameleon, dilaton and symmetron models
\cite{bdlwz2012,bdlwz2013} in general. {\sc ecosmog} has recently been
compared with other codes developed subsequently, including {\sc
  mg-gadget} \cite{mggadget}, Isis \cite{isis} and {\sc mg-enzo}
\cite{mgenzo} and very good agreement was found between all these
codes \cite{code_comparison}.

There are other modified gravity theories, such as the
Dvali-Gabadadze-Porrati \cite{dgp} (DGP) brane-world model, in which
screening is achieved by nonlinear derivative self-couplings of a
scalar field. Well-studied examples include the K-mouflage
\cite{kmouflage1,kmouflage2} and Vainshtein \cite{vainshtein}
mechanisms, the latter being originally studied in massive gravity
theories as a means to suppress the extra helicity modes of massive
gravitons so that GR is recovered in the massless limit. In addition
to the nonlinear massive gravity
\cite{massivegravity1,massivegravity2,cp2012} and braneworld models,
the Vainshtein mechanism is also employed in general setups, such as
the Galileon models \cite{nrt2009,dev2009, Galileon1, Galileon2,
  Galileon3, Galileon4}, which have been the subject of various recent
studies
\cite[e.g.,][]{ck2009,sk2009,gs2010,dt2010,ndt2010,ags2010,bbd2011,al2012,al2012b,blbp2012,nrcpagb2013,blbp2013,blhlbp2014,fkzl2014,fkz2015,ott2013,wjl2013,zwjlw2014,bbll2016,nrabgmb2016}.
      
The first two generations of modified gravity simulation codes
\cite[e.g.,][]{o2008,lkz2008,lz2010,zlk2011} were either not
parallelised or had a uniform resolution across the whole simulation
box, resulting in insufficient resolution and inefficiency. The
current generation of codes, such as {\sc ecosmog}, {\sc mg-gadget},
Isis and {\sc mg-enzo}, are all efficiently parallelised. These codes
solve the nonlinear field equations in modified gravity on meshes (or
their equivalents), and employ the adaptive mesh refinement (AMR)
technique to generate ever finer meshes in high density regions to
increase resolution. However, even with these parallelised codes,
modified gravity simulations currently are still very slow compared to
the fiducial GR case. As we shall discuss below, this is partially due
to the nonlinear nature of the equations to be solved, and partly due
to the specific numerical algorithms used. The greater computational
cost of modified gravity simulations makes it difficult to achieve the
resolution and volume attained in state-of-the-art simulations of
standard gravity.

The coming decade will see a flood of high-precision observational
data from a new generation of cosmological surveys, such as e{\sc
  rosita} \cite{erosita}, the Dark Energy Spectroscopic Instrument
({\sc desi}) \cite{desi,desi1,desi2}, {\sc euclid} \cite{euclid} and
the Large Synoptic Survey Telescope ({\sc lsst}) \cite{lsst}. These
surveys will provide us with golden opportunities to perform
cosmological tests of gravity \citep[see ref.][for a recent
  review]{koyama2016} and seek a better understanding of the origin of
cosmic acceleration. As things stand now, it is the lack of more
powerful simulation methods that limits the accuracy and size that
modified gravity simulations can possibly attain, therefore preventing
us from fully exploiting future observations. This has led to many
attempts to speed up simulations using approximate methods
\cite[e.g.,][]{wf2015,bbl2015}, or develop alternative methods to
predict theoretical quantities
\cite[e.g.,][]{le2012,zhao2014,mpll2015,crll2016}. These alternative
methods are fast substitutes of full simulations and powerful when
quickly exploring a large parameter space is the primary
concern. However, simulations are nevertheless necessary to calibrate
these methods or when better (e.g., \%-level) accuracy is needed, as
well as to study the impact of different theories of gravity on galaxy
formation.

In \cite{bbl2015}, an approximate method to speed up $N$-body
simulations of Vainshtein-type models was presented and shown to
reduce the overhead\footnote{Throughout this paper, the term
  `overhead' is used to refer to the {\it extra} computational time
  (using the same machine and number of cores) involved in running a
  modified gravity simulation compared to standard gravity. For
  example, an overhead of 110\% means that the modified gravity run
  requires $2.1\times$ the CPU time of a $\Lambda$CDM simulation.} of
solving the modified gravity equation to the level of $50\sim100$\%,
with the errors in various cosmological quantities being controlled to
well under $\sim1$\% or smaller (comparable to the discrepancies in
the predictions of different modified gravity simulation codes
\cite{code_comparison}). The same method, however, does not work as
accurately in chameleon-type models (see Appendix~\ref{app:trunc}),
the simulations for which are much more expensive than those for the
Vainshtein-type models. Given that chameleon models are a large class
of modified gravity models that are of interest to the theoretical and
observational community, there is an equally urgent need for fast
simulation methods for them -- this is precisely the purpose of this
paper.

Unlike the truncated simulation method in \cite{bbl2015}, which
artificially suppresses the solver of the modified gravity equation on
higher refinement levels of the AMR meshes, and instead interpolates
the solution on lower (or coarser) refinement levels to find {\it
  approximate} solutions on higher levels, the method proposed here
still solves the {\it full} modified gravity equations on {\it all}
levels. The improved efficiency comes instead from a different way to
discretise the equation on meshes, that makes it less nonlinear and
greatly enhances the rate of convergence of the solution. The new
scheme boosts the performance of the code {\it by a factor of 5} for a
simulation with a periodic box of size $512 \, \rm{Mpc}/h$ and $512^3$
particles, and by a factor of {\it more than 20}, for a higher
resolution setup with a box size of $ 128 \, \rm{Mpc}/h$ and $512^3$
particles.  The method has its own limitation, namely that the
existence of analytical solutions is a particular property of
Hu-Sawicki (HS) $f(R)$ gravity -- as well as a few other examples of
chameleon, symmetron (see Appendix~\ref{app:symmetron}) and dilaton --
models. However, the generic nature of the HS model (in the sense that
with varying parameters it covers a wide range of cosmological
behaviours predicted by various other classes of models) and the lack
of {\it a preferred} fundamental model make a good argument for using
this model as a testbed, given that it is both {\it impossible} and
{\it unnecessary} to study all chameleon-type models using
simulations.

This paper will be arranged as follows. In \S~\ref{sect:model} we
briefly describe the $f(R)$ gravity model and the chameleon screening
mechanism. In \S~\ref{sect:equations} we recap the method currently
employed in $f(R)$ simulations and explain why it is inefficient,
before describing the new method. In \S~\ref{sect:tests} we perform
some tests as validation of this new method. Finally, we discuss and
summarise in \S~\ref{sect:summary}.

Throughout the paper we shall follow the metric convention
$(+,-,-,-)$, and set $c=1$ except in the expressions where $c$ appears
explicitly. Greek indices $\mu, \nu, \cdots$ run over 0, 1, 2, 3. A
subscript $_0$ denotes the present-day value of a quantity.

\section{The Hu-Sawicki $f(R)$ gravity model}

\label{sect:model}

In this section, we briefly review the Hu-Sawicki design
\citep{hs2007} of $f(R)$ gravity, which is currently the most widely
studied one in the literature.

\subsection{The model}

$f(R)$ gravity is obtained by replacing the Ricci scalar $R$ in the
standard Einstein-Hilbert action with an algebraic function of $R$
\cite[see, e.g., ref.][for a review]{sf2010}:
\begin{eqnarray}\label{eq:action}
S &=& \frac{1}{16\pi G}\int{\rm d}^4x\sqrt{|g|}\left[R+f(R)\right] + \int{\rm d}^4x\mathcal{L}_m,
\end{eqnarray}
in which $G$ is Newton's gravitational constant, $g$ the determinant
of the metric tensor $g_{\mu\nu}$ and $\mathcal{L}_m$ is the
Lagrangian density for matter fields (including cold dark matter,
baryons, radiation and neutrinos).
 
The inclusion of $f(R)$ in Eq.~(\ref{eq:action}) changes the Einstein
equation from second order to 4th order in derivatives of the metric
tensor. However, one can straightforwardly rewrite the equation into
two second order ones by defining a new variable, a scalar field,
$f_R\equiv{\rm d}f(R)/{\rm d}R$. These include an equation of motion
that governs the dynamics of the scalar field:
\begin{eqnarray}\label{eq:eom0}
\Box f_R = \frac{1}{3}\left[R-f_RR+2f(R)-8\pi G\rho_m\right] \equiv \frac{\partial V_{\rm eff}\left(f_R\right)}{\partial f_R},\ \ \ \ 
\end{eqnarray}
and a modified Einstein equation:
\begin{eqnarray}\label{eq:einstein}
G_{\mu\nu} &=& 8\pi GT^m_{\mu\nu} + X_{\mu\nu}, 
\end{eqnarray}
with
\begin{eqnarray}
X_{\mu\nu} &\equiv& -f_RR_{\mu\nu} + \left(\frac{1}{2}f-\Box f_R\right)g_{\mu\nu} + \nabla_\mu\nabla_\nu f_R,
\end{eqnarray}
in which $T^m_{\mu\nu}$ is the matter energy-momentum tensor, $\rho_m$
the mass density of non-relativistic matter species, $\nabla_\mu$ the
covariant derivative compatible with metric $g_{\mu\nu}$, $\Box$ the
Laplacian operator, $R_{\mu\nu}$ the Ricci tensor and $G_{\mu\nu}$ the
Einstein tensor.

In the nonlinear regime of structure formation, assuming the
quasistatic \citep{bhl2015} and weak-field approximations,
Eq.~(\ref{eq:einstein}) simplifies to a modified Poisson equation,
\begin{eqnarray}\label{eq:poisson}
\nabla^2\Phi &=& \frac{16\pi G}{3}\delta\rho_m - \frac{1}{6}\delta R\left(f_R\right),
\end{eqnarray}
which relates the gravitational potential $\Phi$ at a given position
to the density ($\delta\rho_m\equiv\rho_m-\bar{\rho}_m$, where a bar
denotes the cosmic mean of a quantity) and curvature ($\delta R\equiv
R-\bar{R}$) at that position.

Similarly, Eq.~(\ref{eq:eom0}) reduces to
\begin{eqnarray}\label{eq:eom}
\nabla^2f_R &=& \frac{1}{3c^2}\left[\delta R\left(f_R\right)-8\pi G\delta\rho_m\right].
\end{eqnarray}
Eqs.~(\ref{eq:poisson}) \& (\ref{eq:eom}) need to be solved in
cosmological simulations for $f(R)$ gravity to predict the modified
gravitational force that is responsible for structure formation. It
can be seen that Eq.~(\ref{eq:eom}) has a similar form to the Poisson
equation, but $\delta R\left(f_R\right)$ is generally a nonlinear
function of $f_R$, and this makes it more difficult to numerically
solve this equation.

Of course, to fully specify a $f(R)$ model one must fix the functional
form $f(R)$. Without the guidance of a fundamental theory, it is not
hard to imagine that there is no unique, or even preferred, way to do
this. However, there are indeed practical considerations that mean
that the functional form cannot be arbitrary either. This is because
the choice of $f(R)$ must serve the purpose that it is originally
designed for: namely, to explain the accelerated cosmic
expansion. Moreover, as we shall see below, the design of $f(R)$ must
ensure that any deviation from GR is suppressed to an insignificant
level in places such as the Solar System, where numerous tests have
confirmed compatibility with GR to high precision. Indeed, it is known
\citep[e.g., refs.][]{bbds2008,whk2012,raveri2014,chl2016} that for
any $f(R)$ model to pass Solar System gravity tests, the background
evolution must be close to (practically indistinguishable from) that
of $\Lambda$CDM.

One functional form of $f(R)$ is proposed by Hu \& Sawicki \cite[HS,
  ref.][]{hs2007}, and has been shown to satisfy these
requirements. It is given as:
\begin{eqnarray}
f(R) &=& -m^2\frac{c_1\left(-R/m^2\right)^n}{c_2\left(-R/m^2\right)^n+1},
\end{eqnarray}
where $n, c_1, c_2$ are dimensionless model parameters, and $m$ is
another model parameter of mass dimension one. To relate $f_R$ to $R$,
we write:
\begin{eqnarray} \label{eq:fr_xi}
f_R = -n\frac{c_1}{c_2^2}\frac{\left(-R/m^2\right)^{(n+1)}}{\left[1+\left(-R/m^2\right)^n\right]^2} \approx -n\frac{c_1}{c_2^2}\left(\frac{m^2}{-R}\right)^{(n+1)},\ \ \ \
\end{eqnarray}
where in the approximation we have used the fact that:
\begin{eqnarray}
-\bar{R} \approx 8\pi G\bar{\rho}_m - 2\bar{f}(R) = 3m^2\left[a^{-3}+\frac{2}{3}\frac{c_1}{c_2}\right].
\end{eqnarray}
Setting $c_1/c_2=6\Omega_\Lambda/\Omega_m$, this model reproduces a
$\Lambda$CDM background expansion history (in which $\Omega_m$,
$\Omega_\Lambda$ are, respectively, the present-day fractional density
of non-relativistic matter and the cosmological constant for this
background). Taking the values of $\Omega_m$, $\Omega_\Lambda$ from
any preferred cosmology, we get $-R\gg m^2$. In what follows, we
define $\xi\equiv c_1/c_2^2$ for brevity, and the background value of
$f_R$ today, $f_{R0}$, which can be obtained from $\xi$.

\subsection{Chameleon screening mechanism}

In the absence of the chameleon screening mechanism
\citep{kw2004,ms2007}, all $f(R)$ gravity models would have been ruled
out by local gravity tests due to the enhanced strength of
gravity. The chameleon screening acts to suppress this enhancement,
thereby allowing $f(R)$ gravity models like HS to pass experimental
constraints on deviations from GR.

The chameleon mechanism works as follows: because the fifth force is
mediated by a scalar field that has a nonzero mass given
by: $$m^2_{f_R}\equiv\frac{\partial^2V_{\rm eff}(f_R)}{\partial^2
  f_R},$$ it is of Yukawa type and proportional to $\exp(-m_{f_R}r)$,
where $r$ is the distance between two test particles. In high matter
density environments, $m_{f_R}$ is heavy and the suppression of the
fifth force becomes very efficient. In reality, this is equivalent to
having $|f_R|\ll1$ in high density regions, which therefore suppresses
the fifth force since it is proportional to $\vec{\nabla}f_R$ (this
can be seen most easily by eliminating the $\delta R$ term in
Eq.~(\ref{eq:poisson}) using Eq.~(\ref{eq:eom}), which shows that
$f_R$ is essentially the potential of the fifth force).

Hence, the functional form of $f(R)$ is critical in determining if the
fifth force can be sufficiently suppressed in dense environments. For
the HS model, it was shown by \cite{hs2007} that $|f_{R0}|<10^{-5}$ is
required to screen the Milky Way. Currently, the strongest constraint
on the value of $|f_{R0}|$ in the HS model comes from the screening of
dwarf galaxies, which requires $|f_{R0}|\lesssim10^{-7}$ (95\% C.L.)
\cite{jvs2013,vcjv2013}. This is a promising way to constrain $f(R)$
gravity, provided astrophysical systematics are well controlled and
the environmental impact on screening is modelled accurately (which
itself will benefit from high resolution simulations).

In cosmology, there are many constraints on $f_{R0}$ as well, and for
recent reviews on this topic the readers are referred to
\cite{lom2014,bs2016}. In \cite{ter2014,wil2015}, X-ray and weak
lensing estimates for the mass of the Coma cluster are combined to
constrain $|f_{R0}|\lesssim10^{-4.2}$ (95\% C.L.). Two of the
strongest constraints to date both come from cluster abundance. In
\cite{cat2014} the authors use X-ray cluster abundance while in
\cite{liu2016} the counts of high-significance weak lensing
convergence peaks are used as a proxy for cluster counts; both studies
find that $|f_{R0}|\lesssim10^{-5.2}$ after carefully accounting for
systematics, even though the data and analyses are very different. In
\cite{cai2015}, it was found that stacked lensing tangential shear of
cosmic voids could potentially place constraints at a similar
level. More recently, a study by \cite{peirone2016}, which uses Planck
Sunyaev-Zel'dovich cluster counts, constrains
$|f_{R0}|\lesssim10^{-5.8}$, although the result is quite sensitive to
the halo mass function used in the analysis. All the constraints are
quoted at 95\% confidence. There are many other cosmological
constraints in the literature but it is beyond the scope of this paper
to mention all of them (some of these studies were carried out by
using linear perturbation theory, which underestimates the
effectiveness of screening and can therefore overestimate the strength
of the constraints on the model -- this is why simulations that fully
capture the nonlinearity of the theories are useful).

\section{$N$-body equations and algorithm}

\label{sect:equations}

In this section, we describe the $N$-body equations in appropriate
code units and their discretised versions that {\sc ecosmog} solves in
simulations.

\subsection{The Newton-Gauss-Seidel relaxation method}
\label{sect:old}

Like its base code {\sc ramses} \cite{ramses}, {\sc ecosmog} adopts
supercomoving coordinates \cite{ms1998} to express the field equations
in terms of dimensionless quantities. The newly defined variables in
these code units are summarised as follows (the tilded quantities are
those in the code units):
\begin{eqnarray}\label{eq:code_unit}
\tilde{x}\ =\ \frac{x}{B},\ \ \ \tilde{\rho}\ =\ \frac{\rho a^3}{\rho_c\Omega_m},\ \ \ \tilde{v}\ =\ \frac{av}{BH_0},\nonumber\\
\tilde{\Phi}\ =\ \frac{a^2\Phi}{(BH_0)^2},\ \ \ d\tilde{t}\ =\ H_0\frac{dt}{a^2},\ \ \ \tilde{f}_R\ =\ \frac{c^2a^2f_R}{(BH_0)^2}, \nonumber
\end{eqnarray}
in which $x$ is the comoving coordinate, $\rho_c$ is the critical
density today, and $v$ the particle velocity. In addition, $B$ is the
comoving size of the simulation box in units of $\rm{Mpc}/h$ and $H_0$
is the Hubble expansion rate today in units of $100h$~km/s/Mpc. Note
that with these conventions, the mean matter density is
$\tilde{\bar{\rho}}=1$. All the tilded quantities are dimensionless.

In terms of these variables, the Poisson and scalar field equations
(Eqs.~ \ref{eq:poisson} \& \ref{eq:eom}) in the HS model can be
rewritten as:
\begin{widetext}
\begin{eqnarray}
\label{eq:poisson_code} \tilde{\nabla}^2\tilde{\Phi} &=& 2\Omega_m(\tilde{\rho}-1) - \frac{1}{6}\Omega_ma^4\left[\left(-\frac{na^2\xi}{\tilde{f}_R}\right)^{\frac{1}{n+1}}-3\left(a^{-3}+4\frac{\Omega_\Lambda}{\Omega_m}\right)\right],\\
\label{eq:eom_code} \tilde{\nabla}^2\tilde{f}_R &=& -\frac{1}{\tilde{c}^2}\Omega_ma(\tilde{\rho}-1) + \frac{1}{3\tilde{c}^2}\Omega_ma^4\left[\left(-\frac{na^2\xi}{\tilde{f}_R}\right)^{\frac{1}{n+1}}-3\left(a^{-3}+4\frac{\Omega_\Lambda}{\Omega_m}\right)\right],
\end{eqnarray}
\end{widetext}
in which we have used the relation $m^2 = \Omega_mH_0^2$, and defined
$\tilde{c}=c/(BH_0)$, which is the speed of light in code units.

In principle, Eqs.~(\ref{eq:poisson_code}) \& (\ref{eq:eom_code}) can
be directly discretised on a mesh and can then be solved
numerically. For chameleon-type models, however, there is a further
subtlety: namely, the value of $-\tilde{f}_R$ is very small at early
times and in high density regions. This property is desirable in order
that the model can pass Solar System tests of gravity by virtue of the
chameleon mechanism, but it also poses a challenge when numerically
solving Eq.~(\ref{eq:eom_code}). In the relaxation method that is
employed to solve the discrete version of this equation,
$-\tilde{f}_R$ in each mesh cell gets updated until the solution is
close enough to the true value (more details below). This updating
procedure is a numerical approximation, and it is possible that
$-\tilde{f}_R$ can acquire negative numerical values in some cells as
a result. Taking the case of the HS $n=1$ model as an example: the
quantity $(-\tilde{f}_R)^{\frac{1}{n+1}}$ is not physically defined if
$-\tilde{f}_R<0$, and the code then runs into trouble.

To overcome this numerical issue, in \cite{o2008} Oyaizu proposes to
replace $-\tilde{f}_R$ with $\exp(u)$ in Eq.~(\ref{eq:eom_code}). As
$\exp(u)$ can only be positive, this guarantees that the nonphysical
situation described above will never appear. This change of variable
has since then been used in all simulation codes of chameleon models
to our knowledge
\cite{olh2008,sloh2009,lz2009,zmlhf2010,lz2010,zlk2011,ecosmog,mggadget,isis,mgenzo}.

In terms of this new variable, Eq.~(\ref{eq:eom_code}) can be
discretised as:
\begin{widetext}
\begin{eqnarray}\label{eq:eom_discrete_old}
\frac{1}{h^2}\left[b_{i+\frac{1}{2},j,k}u_{i+1,j,k} - u_{i,j,k}\left(b_{i+\frac{1}{2},j,k}+b_{i-\frac{1}{2},j,k}\right) + b_{i-\frac{1}{2},j,k}u_{i-1,j,k}\right] + \nonumber\\
\frac{1}{h^2}\left[b_{i,j+\frac{1}{2},k}u_{i,j+1,k} - u_{i,j,k}\left(b_{i,j+\frac{1}{2},k}+b_{i,j-\frac{1}{2},k}\right) + b_{i,j-\frac{1}{2},k}u_{i,j-1,k}\right] + \nonumber\\
\frac{1}{h^2}\left[b_{i,j,k+\frac{1}{2}}u_{i,j,k+1} - u_{i,j,k}\left(b_{i,j,k+\frac{1}{2}}+b_{i,j,k-\frac{1}{2}}\right) + b_{i,j,k-\frac{1}{2}}u_{i,j,k-1}\right] + \nonumber\\
\frac{1}{3\tilde{c}^2}\Omega_ma^4\left(na^2\xi\right)^{\frac{1}{n+1}}\exp\left[-\frac{u_{i,j,k}}{n+1}\right] - \frac{1}{\tilde{c}^2}\Omega_ma(\rho_{i,j,k}-1) - \frac{1}{\tilde{c}^2}\Omega_ma^4\left(a^{-3}+4\frac{\Omega_\Lambda}{\Omega_m}\right) &=& 0,
\end{eqnarray}
\end{widetext}
in which we have used the second order finite difference scheme to
calculate $\tilde{\nabla}^2\left(-\tilde{f}_R\right)$. Taking the
second order derivative with respect to the $x$ coordinate as an
example, this scheme gives:
\begin{eqnarray}
\frac{\partial^2\phi}{\partial x^2} \rightarrow \frac{1}{h^2}\left(\phi_{i+1,j,k}-2\phi_{i,j,k}+\phi_{i-1,j,k}\right),\nonumber
\end{eqnarray}
where $h$ is the size of the mesh cell and the subscript $_{i,j,k}$
refers to the cell that is $i$-th in the $x$ direction, $j$-th in the
$y$ direction and $k$-th in the $z$ direction. Note that the discrete
Laplacian in Eq.~(\ref{eq:eom_discrete_old}) looks slightly more
complicated because
$\tilde{\nabla}^2\exp(u)\equiv\tilde{\nabla}\cdot\left(e^u\tilde{\nabla}u\right)$,
and we have defined $b\equiv\exp(u)$ such that:
\begin{eqnarray}
b_{i+\frac{1}{2},j,k} &\equiv& \frac{1}{2}\left[\exp\left(u_{i+1,j,k}\right)+\exp\left(u_{i,j,k}\right)\right],\nonumber\\
b_{i-\frac{1}{2},j,k} &\equiv& \frac{1}{2}\left[\exp\left(u_{i-1,j,k}\right)+\exp\left(u_{i,j,k}\right)\right].\nonumber
\end{eqnarray}
Defining the left-hand side of Eq.~(\ref{eq:eom_discrete_old}) as the
operator $\mathcal{L}^h$, where a superscript $^h$ is used to denote
that the equation is discretised on a mesh with cell size $h$, the
equation can be written symbolically as:
\begin{eqnarray}
\mathcal{L}^h(u_{i,j,k}) &=& 0.
\end{eqnarray}
This is a nonlinear equation for $u_{i,j,k}$, and the most commonly
used method to solve it is relaxation, which begins with some initial
guesses of $u_{i,j,k}$ (for all mesh cells) and iteratively improves
the old guess to get a new guess according to the Newton-Raphson
method (same as the one used for solving nonlinear algebraic
equations):
\begin{eqnarray}\label{eq:relaxation_update}
u^{h,{\rm new}}_{i,j,k} &=& u^{h,{\rm old}}_{i,j,k} - \frac{\mathcal{L}^{h}\left(u^{h,{\rm old}}_{i,j,k}\right)}{\frac{\partial\mathcal{L}^h\left(u^{h,{\rm old}}_{i,j,k}\right)}{\partial u^h_{i,j,k}}},
\end{eqnarray}
until $u_{i,j,k}$ (for all mesh cells) is close enough to the true
solution or, equivalently, some all-mesh average of
$\mathcal{L}^h\left(u_{i,j,k}\right)$ gets close enough to zero. A
widely used definition of this all-mesh average (the so-called
residual) is given by:
\begin{eqnarray}
{\rm Residual} &\equiv& \left[\sum_{i,j,k}\left[\mathcal{L}^h\left(u_{i,j,k}\right)\right]^2\right]^{1/2},
\end{eqnarray}
where the summation is performed over all mesh cells on a given
refinement level.

The implementation of this method is fairly straightforward in
principle, but in practice there are a number of subtleties that need
to be taken into account. For example, refined meshes usually have
irregular shapes and their boundary conditions should be carefully set
up by interpolating the values of $u$ from coarser levels. The
relaxation method is also notoriously slow to converge (convergence
here meaning that the residual becomes smaller than some pre-fixed
threshold) if it is only done on a fixed level, and in practice the
so-called multigrid method is commonly used to remedy this
\citep{brandt1977}. This consists of moving the equation to coarser
meshes, solving it there, and then using the coarse-mesh solutions to
correct the fine-mesh one. These subtleties have been discussed in
detail in the literature; as they are not the main concern of this
paper, we refer interested readers to, e.g., \cite{ecosmog}, for a
more elaborate description.

Although the multigrid method improves convergence in general, the
rate of convergence is still very slow in $f(R)$ simulations, and the
relaxation is some times unstable and diverges. One way to improve
both the rate of convergence and the stability of the
Newton-Gauss-Seidel relaxation method is to impose the condition:
\begin{eqnarray}
\left|\mathcal{L}^h (u_{i,j,k}^{h,\rm{new}})\right| < \left|\mathcal{L}^h (u_{i,j,k}^{h,\rm{old}}) \right|, \nonumber
\end{eqnarray}
i.e., requiring that the residual after the new iteration gets
monotonically smaller than in the previous one. When the condition is
not met, we retain the value of the scalar field from the previous
step ($u_{i,j,k}^{h,\rm{old}}$). While satisfying this condition can
be costly on each step, the overall efficiency of the code can be
significantly increased by the improved numerical stability and
reduced number of iterations required to reach convergence.

Finally, a similar discretisation can be done for the modified Poisson
equation:
\begin{widetext}
\begin{eqnarray}\label{eq:poisson_discrete}
&&\frac{1}{h^2}\Big(\Phi_{i+1,j,k}+\Phi_{i-1,j,k}+\Phi_{i,j+1,k}+\Phi_{i,j-1,k}+\Phi_{i,j,k+1}+\Phi_{i,j,k-1}-6\Phi_{i,j,k}\Big)\nonumber\\ 
&=& 2\Omega_ma(\rho_{i,j,k}-1) - \frac{1}{6}\Omega_ma^4\left[\left(na^2\xi\right)^{\frac{1}{n+1}}\exp\left(-\frac{u_{i,j,k}}{n+1}\right) - 3\left(a^{-3}+4\frac{\Omega_\Lambda}{\Omega_m}\right)\right].
\end{eqnarray}
\end{widetext}
This equation is solved after Eq.~(\ref{eq:eom_discrete_old}), by
which time $u_{i,j,k}$ is already known. As a result, this is a linear
equation for $\Phi_{i,j,k}$ which is easier to solve than
Eq.~(\ref{eq:eom_discrete_old}), and we shall not discuss it further
here. Structurally, Eq.~(\ref{eq:poisson_discrete}) is the same as the
Poisson equation for standard gravity (with a modified source term);
hence, one may simply use the standard {\sc ramses} implementations
for solving the Poisson equation.

\subsection{The new method}
\label{sect:new}

The discretisation used in the scalar field equation
(Eq.~\ref{eq:eom_discrete_old}) has a number of drawbacks:

\begin{itemize}

\item Depending on the value of $\xi$, the original scalar field
  equation can be very nonlinear (when $\xi$ is small, the term
  involving $\left(-\tilde{f}_R \right)^{\frac{1}{n+1}}$ is large and
  non-negligible, c.f. Eq.~\ref{eq:fr_xi}) or close to linear (when
  $\xi$ is large, that term is small and negligible so that the
  equation becomes nearly linear in $\tilde{f}_R$)
   \footnote{Note that, on first glance at
     Eq.~(\ref{eq:eom_discrete_old}), this may appear
     counter-intuitive. This dependence of the degree of linearity of
     Eq.~(\ref{eq:eom_discrete_old}) on the size of $\xi$ can be
     explained by the fact that as $\xi$ becomes smaller, the value of
     $\tilde{f}_{R}$ also becomes smaller (c.f. Eq.~\ref{eq:fr_xi}),
     making Eq.~(\ref{eq:eom_discrete_old}) on the whole more
     nonlinear. The converse is true when $\xi$ is large.}. In the
   former case, introducing the new variable $u=\log(-\tilde{f}_R)$
   makes the equation even more nonlinear; in the latter case, it
   nonlinearises an almost linear equation. The high nonlinearity
   makes the relaxation method very slow to converge, which is why
   simulations of $f(R)$ gravity are generally much more costly than
   $\Lambda$CDM simulations with the same specifications. Indeed, even
   with parallelised codes such as {\sc ecosmog}, {\sc mg-gadget},
   Isis and {\sc mg-enzo} (Zhao et al. in prep.), very large-sized and
   high resolution $f(R)$ simulations are currently still difficult to
   run, and this situation needs to be improved if we want to compare
   future survey data to theoretical predictions to perform accurate
   tests of modified gravity.

\item As we have already seen above, the discrete Laplacian
  $\tilde{\nabla}^2e^u$ is more complicated than the simple
  discretisation of $\tilde{\nabla}^2\tilde{\Phi}$, resulting in a
  more complex equation that needs to be solved.

\item The code ends up with a lot of $\exp$ and $\log$
  operations. This is not optimal from a practical viewpoint, because
  the cost of these operations is generally much higher than that of
  simple arithmetic ones, such as summation, subtraction and
  multiplication.

\end{itemize}

The method described here alleviates the nonlinearity problem by
defining a new variable $u=\left(-\tilde{f}_R\right)^{1/2}$, so that
the scalar field equation for the HS model with $n=1$ (the most widely
studied $f(R)$ model in the literature) becomes a simple cubic
equation in $u$, which can be solved analytically instead of resorting
to the approximation in Eq.~(\ref{eq:relaxation_update}):
\begin{eqnarray}\label{eq:cubic_equation}
u^3_{i,j,k} + pu_{i,j,k} + q= 0,
\end{eqnarray}
where: 
\begin{widetext}
\begin{eqnarray} \label{eq:new_discrete}
p &\equiv& \frac{h^2}{6\tilde{c}^2}\Omega_ma\tilde{\rho}_{i,j,k} + \frac{2h^2}{3\tilde{c}^2}\Omega_\Lambda a^4 - \frac{1}{6}\left(u^2_{i+1,j,k}+u^2_{i-1,j,k}+u^2_{i,j+1,k}+u^2_{i,j-1,k}+u^2_{i,j,k+1}+u^2_{i,j,k-1}\right),\\
q &\equiv& - \frac{h^2}{18\tilde{c}^2}\Omega_ma^4\xi^{1/2}.
\end{eqnarray}
\end{widetext}
Note that here we focus on the case of $n=1$; other cases will be
discussed later.

While Eq.~(\ref{eq:cubic_equation}) can be solved analytically (and
therefore accurately), it has three branches of solutions and,
depending on the numerical values of $p$ and $q$, all these branches
can be real. Therefore, extra care has to be taken to make sure that
the correct branch of solutions is chosen. For this, let us define:
\begin{eqnarray}
\Delta_0 &\equiv& -3p,\nonumber\\
\Delta_1 &\equiv& 27q.
\end{eqnarray}
As $q<0$ is a constant in a given time step of the simulation, we see
that $\Delta_1<0$.

The case $p>0$ can occur in high density regions where $u>0$ is small
(and $u^2$ smaller still) because of the chameleon screening. In these
cases, $\Delta_0<0$ and thus $\Delta_1^2-4\Delta_0^3>0$. The cubic
equation then admits only one real solution, which must be the one we
choose:
\begin{eqnarray}\label{eq:cubic_solution_1}
u_{i,j,k} &=& -\frac{1}{3}\left(C+\frac{\Delta_0}{C}\right)
\end{eqnarray}
with
\begin{eqnarray} \label{eq:C_equation}
C &\equiv& \left[\frac{1}{2}\left(\Delta_1+\left(\Delta_1^2-4\Delta_0^3\right)^{1/2}\right)\right]^{1/3}.
\end{eqnarray}
Note that Eq.~(\ref{eq:C_equation}) implies that $C=0$ only when
$\Delta_0 = p = 0$. This ensures that for the $p>0$ case, $C \neq 0$
in Eq.~(\ref{eq:cubic_solution_1}), and the solution is never
undefined.

In the case of $p=0$, the solution is simply:
\begin{eqnarray}\label{eq:cubic_solution_2}
u_{i,j,k} &=& (-q)^{1/3}.
\end{eqnarray}

$p<0$ can occur for density peaks in an overall low density region
(where $u$ and hence $u^2$ can be large).  $\Delta_1^2-4\Delta_0^3$
can then take either positive or negative values. In the former case,
the solution in Eq.~(\ref{eq:cubic_solution_1}) still holds, while in
the latter case the equation has three real solutions:
\begin{eqnarray}
u_{i,j,k} &=& -\frac{2}{3}\Delta_0^{1/2}\cos\left[\frac{1}{3}\Theta+\frac{2}{3}j\pi\right],
\end{eqnarray}
where $j=0,1,2$ and
$\cos\Theta\equiv\Delta_1/\left(2\Delta_0^{3/2}\right)$. It is
straightforward to decide which branch we should take: as
$\Delta_1<0$, we have $\cos\Theta<0$ and so $\Theta\in(\pi/2,\pi)$.
Given that we require $u_{i,j,k}$ to be positive-definite:

\begin{itemize}

\item If $j=0$, $u_{i,j,k}\sim-\cos\Big(\frac{1}{3} \Theta\Big)<0$ and
  is unphysical;

\item If $j=1$,
  $u_{i,j,k}\sim-\cos\Big(\frac{1}{3}\Theta+\frac{2}{3}\pi\Big)>0$ and
  is physical;

\item If $j=2$,
  $u_{i,j,k}\sim-\cos\Big(\frac{1}{3}\Theta-\frac{2}{3}\pi\Big)<0$ and
  is unphysical.

\end{itemize}

This new method has a few interesting features:

\begin{itemize}

\item The discrete equation to be solved is significantly simpler. In
  particular, $q$ is the same in all cells, so it only needs to be
  calculated once for a given time step and on a given mesh refinement
  level.

\item There is a substantial reduction of costly computer operations
  as we get rid of operations. Some $\cos$ and $\cos^{-1}$ operations
  are introduced, but they will not be executed for all cells
  (depending on which branch of solutions we take); even for cells in
  which they need to be executed, they are only executed once. In the
  old method, $\exp$ is executed on both the cell and its neighbours.

\item The cubic equation is solved analytically and a physical
  solution always exists. The variable redefinition in the old method,
  $\tilde{f}_R = \exp(u)$, was chosen so as to the avoid the
  unphysical solution $-\tilde{f}_R < 0$; the new method avoids this
  situation automatically by selecting the physical solution
  $u=\left(-\tilde{f}_R\right)^{1/2} > 0$ analytically. As a result,
  we can expect this new method to be both more stable (i.e., not
  suffering from catastrophic divergences due to numerics) and more
  efficient (i.e., the solution to Eq.~(\ref{eq:cubic_equation}) is
  exact for each Gauss-Seidel iteration, while
  Eq.~(\ref{eq:relaxation_update}) implicitly uses the approximate
  Newton-Raphson method and may need to be executed many times to
  arrive at what the new method achieves in one go).

\end{itemize}

Note that this new method does not really get rid of Gauss-Seidel
relaxation, because the quantity $p$ in Eq.~(\ref{eq:cubic_equation})
depends on the values of the scalar field in (the 6 direct)
neighbouring cells, which are not accurate values but temporary
guesses. It therefore still needs to do the Gauss-Seidel iterations
(we use the standard red-black chessboard scheme here). What it does
get rid of is the `Newton-Raphson' part
[Eq.~(\ref{eq:relaxation_update})] of the Newton-Gauss-Seidel (or
nonlinear GS, or NGS) relaxation which updates the old guesses using a
linear approximation of the full nonlinear equation. The speedup is
also largely assisted by the simplicity of Eq.~(\ref{eq:new_discrete})
compared to Eq.~(\ref{eq:eom_discrete_old}), which comes about due to
the new variable redefinition. Therefore, while Gauss-Seidel
iterations are still required, the savings using the new method can
still be significant.

\section{Tests and Simulations of the New Method}

\label{sect:tests}

In this section we present the results of several test runs of the new
{\sc ecosmog} code. In what follows, we will only consider the F6
model of $f(R)$ gravity, in which the present-day value of the scalar
field is given by $\left| \bar{f}_{R0} \right| = 10^{-6}$. In this
model, the chameleon screening is particularly efficient, meaning that
deviations from GR are very small. To capture the effects of
screening, accurately solving the nonlinear scalar field equations is
therefore necessary.

We have simulated the F6 model at three resolution levels: `Low res',
`Medium res' and `High res' (the box size and number of particles used
in each of these runs are summarised in
Table~\ref{tab:SimDetails}). In each case, we have also run a
$\Lambda$CDM simulation starting from the same initial conditions. The
mesh refinement criteria used for the `High res' simulation allows us
to resolve small scales comparable to those in the Millennium
simulation \citep{millennium}. While the `Low' and `High res' runs use
Planck 2015 \citep{planck2015} cosmological parameters (with $\Omega_m
= 0.308, \Omega_\Lambda = 0.692, h = 0.6781, \sigma_8 = 0.8149$), the
cosmological parameters for the `Medium res' run are obtained from
WMAP-7 \citep{wmap7} data (with $\Omega_m = 0.271, \Omega_\Lambda =
0.729, h = 0.704, \sigma_8 = 0.8092$).

\begin{table} \centering
\begin{tabular}{c c c c c c c}
\hline
\hline
Name &   Model &  $B$  &  $N_p$ & Speed up & Overhead \\
     &         & $\left[ \rm{Mpc}/h \right]$ &   &          & (new method)\\
\hline
Low res & $\Lambda$CDM, F6 &  $512$ & $512^3$ & $5 \times$       & 110\% \\
Medium res & $\Lambda$CDM, F6 & $250$ & $512^3$ & $15 \times$    & 180\% \\
High res &  $\Lambda$CDM, F6 & $128$ & $512^3$ &  $> 20 \times$   & 190\% \\
\hline
\end{tabular}
\caption{Details of the simulations performed in this work. The
  columns $B$ and $N_p$, respectively, refer to the comoving box size
  and number of particles in each of these runs. The starting redshift
  in all simulations was $z_{\rm ini} = 49$. The second last column
  summarises the factor by which the new method is faster than the old
  one in each case. Note that the $> 20 \times$ speedup for the `High
  res' simulation is an estimate - we have not run an F6 simulation at
  this resolution using the old method. The last column shows the
  percentage overhead of the F6 simulations using the new method
  compared to $\Lambda$CDM. The level of speedup that can be achieved
  in the F6 simulations depends on the convergence criteria used: in
  all cases, convergence is considered as achieved when the residual
  is $< 10^{-8}$ on the domain level, and $< 10^{-7}$ on the fine
  levels.}
\label{tab:SimDetails}
\end{table}

\begin{figure}
\centering
\includegraphics[width=\columnwidth]{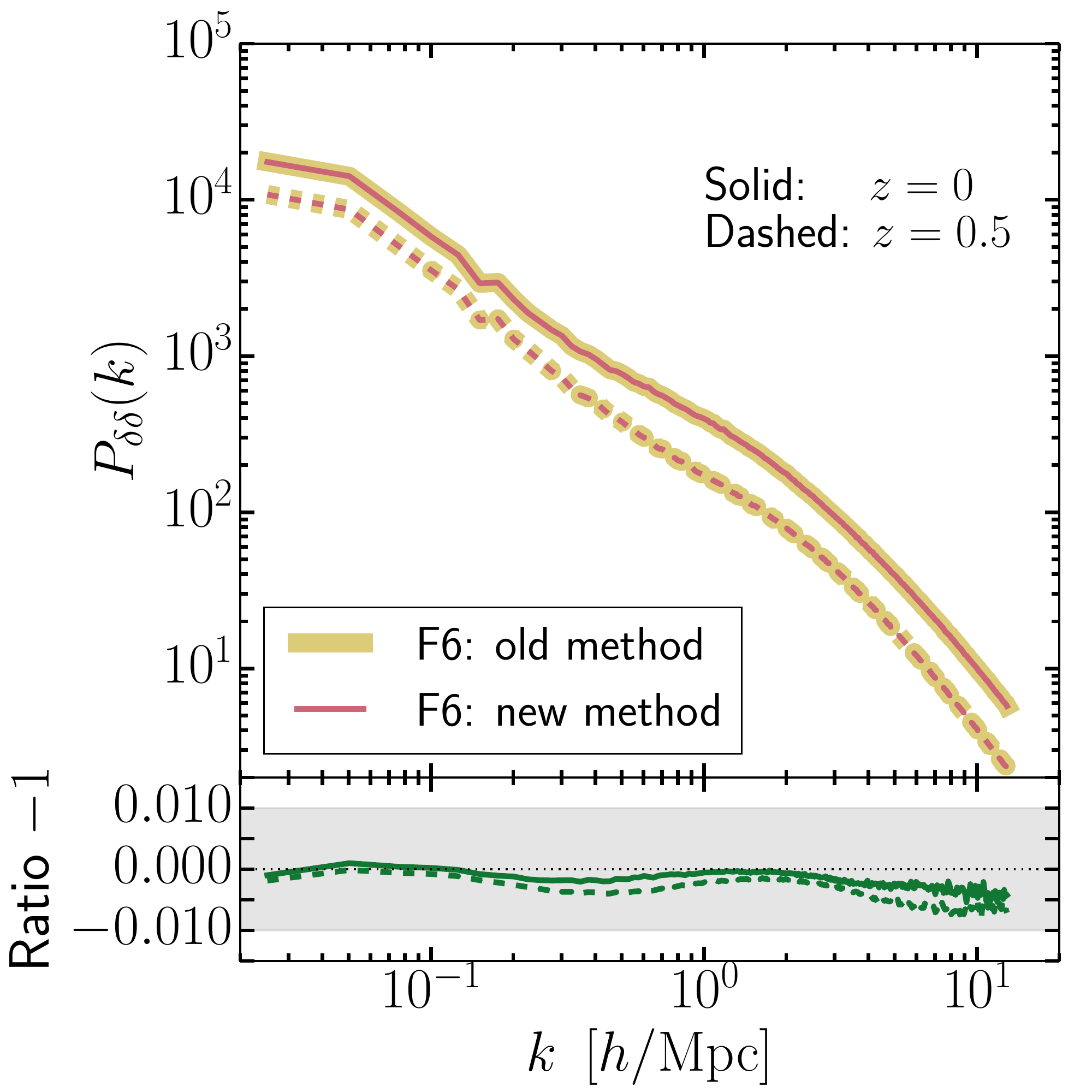}
\caption{{\it Top panel}: Comparison of the nonlinear matter power
  spectra at $z = 0, 0.5$ for the F6 model using the old method
  (yellow, \S~\ref{sect:old}) and the new method (red,
  \S~\ref{sect:new}) for solving the scalar field equations of
  motion. The results shown are for the `Medium res' simulation. {\it
    Bottom panel}: Ratio of the power spectra corresponding to the
  upper panel. The shaded grey band represents a 1\% error region.}
\label{fig:method_compare}
\end{figure}

In Fig.~\ref{fig:method_compare}, we compare the nonlinear matter
power spectrum, $P_{\delta \delta}(k)$, from the `Medium res'
simulations using the old and new methods. $P_{\delta \delta}(k)$ was
computed using the publicly-available {\sc powmes} code
\citep{powmes}. The solid and dashed curves are $P_{\delta \delta}(k)$
computed at $z=0$ and $z=0.5$, respectively, for F6. The results of
the two methods are indistinguishable at both redshifts, and this is
quantified more clearly in the bottom panel of
Fig.~\ref{fig:method_compare}, which shows the relative difference
between the old and new methods. The shaded grey band in this panel
represents a 1\% error around zero; clearly, the new and old methods
agree to well below 1\% at all scales resolved in the simulation. The
same is true even at higher redshift ($z=1, 2$, not shown). We have
checked that the agreement also holds in the case of the velocity
divergence power spectrum, $P_{\theta \theta}(k)$, which, being just
the first integral of the gravitational acceleration, would be more
sensitive to differences in the gravitational forces between the two
methods. Agreement for $P_{\theta \theta}(k)$, which is calculated in
a volume-weighted way, shows that the two methods agree well even in
regions of the cosmic web that are not mass-dominated. This is not
unexpected: after all, the new method solves the same equation of
motion, without needing to use the approximate and inefficient
Newton-Raphson scheme. As a consequence, the simulation is now {\it
  significantly} faster than before: the new method boosts the speed
of the F6 calculation by a factor of 15 relative to the old
implementation in {\sc ecosmog} (see the last column of
Table~\ref{tab:SimDetails}).

\begin{figure}
\centering
\includegraphics[width=\columnwidth]{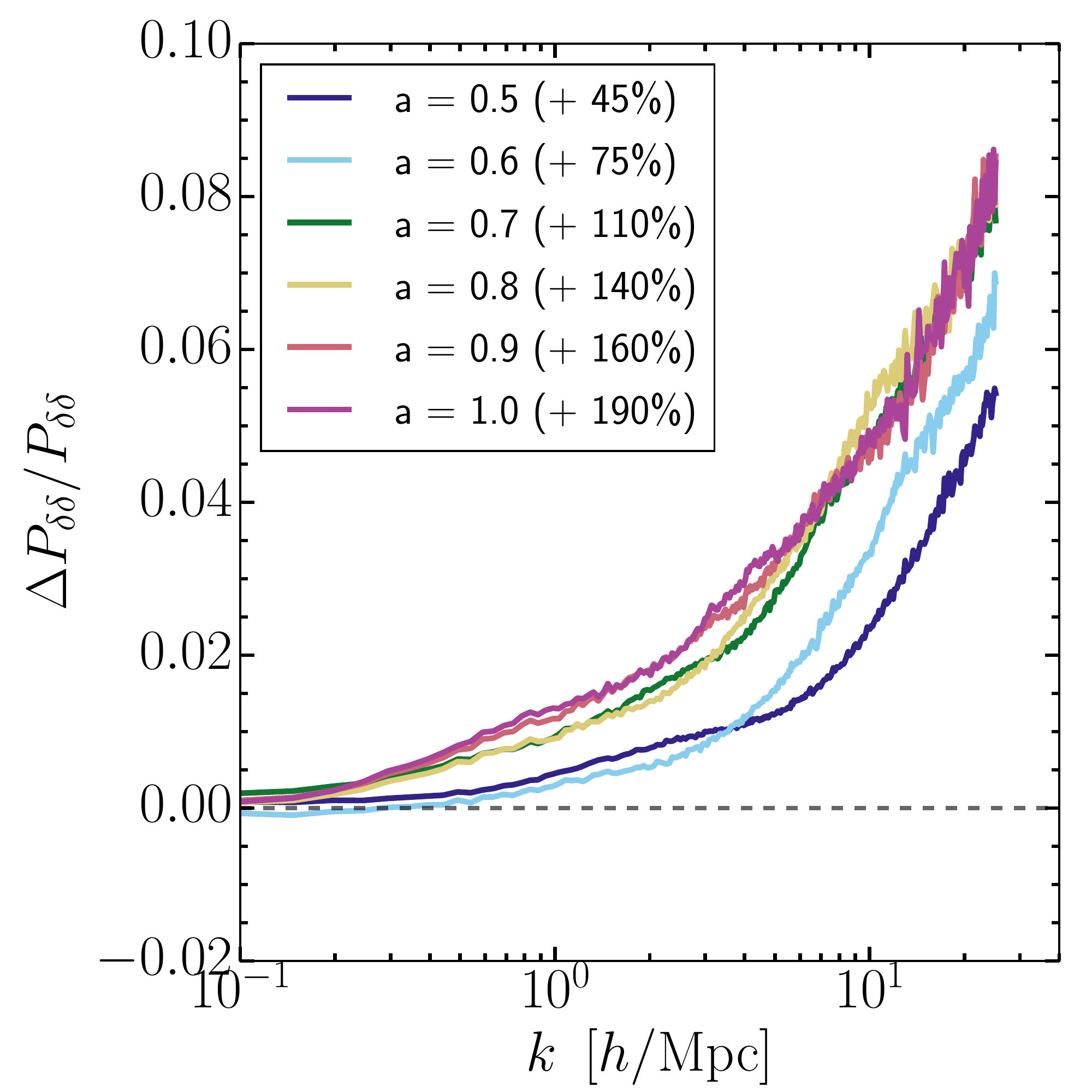}
\caption{Enhancement of the F6 matter power spectrum relative to
  $\Lambda$CDM for the `High res' simulation ($B = 128 \, \rm{Mpc}/h$,
  $N_p = 512^3$). The different coloured curves show the relative
  difference at different scale factors, as indicated in the
  legend. Alongside the legend labels, we also note the percentage
  overhead associated with the F6 run compared to the $\Lambda$CDM run
  at the same scale factor.}
\label{fig:HR_overhead}
\end{figure}

Two-point statistics such as the power spectrum offer a complete
description of clustering properties only for Gaussian
fields. Gravitational instability theory predicts that the nonlinear
evolution induced by gravity drives away the PDF of these fields from
Gaussianity at late times and small scales \cite[see e.g.,
  refs.][]{jbc1993,bernardeau1994}. This is reflected in growing
skewness and kurtosis of cosmic density and velocity fields. $f(R)$
theories show systematic deviations from $\Lambda$CDM for these
statistics, and these can therefore be used as a test of the theory
\citep{hlfc2013}. We have computed PDFs and their higher-order moments
for the density and velocity divergence fields to test how well the
old and new methods agree beyond simple two-point statistics. We find
that the differences are very small and comparable to the differences
seen in the $P(k)$. As an additional test, we have also computed the
Fourier mode decoherence functions \citep{sydhf1992,cc2002}, defined
as Pearson correlation coefficients for the Fourier modes of the two
fields:
\begin{eqnarray}
  C(k)\equiv \frac{\langle f_1 f_2^*\rangle}{\langle
    f_1^2\rangle^{1/2} \langle f_2^2\rangle^{1/2}}, \nonumber
\end{eqnarray}
where $f_1$ and $f_2$ are the density or velocity divergence fields
for the $f(R)$ runs computed using the two methods. $C(k) = 1$ when
both fields being compared have Fourier modes at given $k$ that
correspond exactly. The density and velocity divergence fields for the
F6 runs using the two methods take $C(k)=1$ for almost the entire
range of $k$, up until the Nyquist limit of the simulations. These
tests reassure us that the density and velocity fields produced by the
old and the new method are, for all practical purposes,
indistinguishable.

Results from the `High res' simulations are shown in
Fig.~\ref{fig:HR_overhead}, where we plot the relative difference in
$P_{\delta \delta}(k)$ of F6 with $\Lambda$CDM -- only results using
the new method are shown. Curves of different colours represent the
relative difference at different scale factors, as labelled in the
legend. The legend labels also list the percentage overhead involved
in the F6 simulation compared to the $\Lambda$CDM run at the same
scale factor. With the new method, the F6 simulation is now only $\sim
45\%$ slower than the $\Lambda$CDM run at $a=0.5 \, (z=1)$, and only
$\sim 190\%$ slower at the final time. Compared to F6 simulations with
comparable resolution using the old method \citep[e.g.,
  ref.][]{liminality}, the new implementation is estimated to be {\it
  more than $20 \times$ faster}.

The degree to which the new method improves the efficiency of {\sc
  ecosmog} over the old one depends on resolution. Indeed, in going
from the `Low res' to the `High res' simulations, the gain in
performance increases from a factor of 5 to a factor of over 20 (the
overhead increases considerably with resolution in the old
method). The improved efficiency of the numerical algorithm will
enable us to run simulations of chameleon models that would previously
have been computationally very expensive to perform. Future
applications of the method could include running hydrodynamical
simulations (where high resolution is required to follow accurately
the hydrodynamics and to resolve spatial scales important for star
formation and feedback), and running large numbers of low resolution
volumes to estimate the covariance matrix in non-standard gravity.

\section{Summary and Discussions}

\label{sect:summary}

Modified gravity models are an umbrella group of theories seeking to
explain the apparent accelerated cosmic expansion by assuming
modifications to the Einsteinian gravitational law on cosmological
scales. Usually, such modifications must be small in high density
environments in which gravity is known to be accurately described by
GR, and this can be achieved by screening mechanisms, resulting in
highly nonlinear field equations. Studying the cosmological
implications of these theories and observational constraints on them
is an active research topic in cosmology, but the nonlinear nature of
these theories means that one has to resort to numerical simulations,
which can be prohibitively slow. This has, up until now, limited the
scope of accurately testing gravity using precision observational
data.

In this paper, we proposed and demonstrated the power of a new and
more efficient method to solve the nonlinear field equation in one of
the most popular modified gravity models -- the Hu-Sawicki variant of
$f(R)$ gravity. The current method used to simulate this model is slow
mainly because of a variable redefinition aimed at making the
relaxation algorithm numerically stable, but has the negative side
effect of making the discrete equation even more nonlinear and,
therefore, harder to converge. As a result, modified gravity
simulations which match the size {\it and} resolution of the
state-of-the-art $\Lambda$CDM $N$-body or hydrodynamical simulations
have thus far been beyond reach \citep[but see
  refs.][]{hlmw2015,asp2016}.

The new method avoids the specific variable redefinition used in the
old method, and therefore does not further increase the nonlinearity
of the discrete equation to be solved. More importantly, it enables
the discrete equation to be written in a form that is {\it
  analytically} solvable at each Gauss-Seidel iteration. This is what
ultimately makes the method efficient: compare solving a highly
nonlinear algebraic equation analytically and solving the same
equation using the Newton-Raphson iteration method
(Eq.~\ref{eq:relaxation_update}), and it is clear that the latter is
generally much more inefficient.

We have performed test simulations using the new method, and confirmed
that it is indeed very efficient. The working model for the tests is
the F6 variant of Hu-Sawicki $f(R)$ gravity. The chameleon screening
is very efficient in F6, and it is therefore important that the
nonlinear scalar field equations are solved accurately. In
Fig.~\ref{fig:method_compare}, we have confirmed that the new and old
methods agree at the sub-percent level when comparing the nonlinear
matter power spectrum, $P_{\delta \delta}(k)$. The good agreement
continues to hold at higher redshift, as well as for the velocity
divergence power spectra, $P_{\theta \theta}(k)$. Next, in
Fig.~\ref{fig:HR_overhead}, we presented results from our `High res'
simulations, which are comparable in resolution to the Millennium
simulation. The total overhead in the F6 simulation is $\sim 190\%$
compared to the equivalent $\Lambda$CDM run; this represents a boost
in performance of $> 20\times$ compared to an F6 simulation of similar
resolution using the old method.

The improved performance of the new simulation algorithm compared to
the old one serves to highlight the importance of the way in which one
discretises partial differential equations for the efficiency of
numerically solving them. This is particularly true for highly
nonlinear equations, such as those encountered in many modified
gravity theories. Our work highlights the following:

(1) There is not a single way of discretisation, and this usually
depends on the specific equations to be solved. In general, the
discretisation should be chosen to preserve the original degree of
nonlinearity of the equation as much as possible, and avoid further
nonlinearising the equation.

(2) Where possible, exact solutions to the nonlinear discrete equation
should be used instead of the approximate solution in
Eq.~(\ref{eq:relaxation_update}). The latter, despite being commonly
used in relaxation solutions to nonlinear differential equation, is a
second option only for cases where
$\mathcal{L}^h\left(u_{i,j,k}\right)=0$ has no analytical solution in
general.

The same observations and conclusions apply to other classes of
partial differential equations, such as those involving higher order
powers of the derivatives of the scalar field (e.g.,
$\left(\nabla^2\varphi\right)^2$,
$\nabla^i\nabla^j\varphi\nabla_i\nabla_j\varphi$,
$\left(\nabla^2\varphi\right)^3$,
$\nabla^i\nabla_j\varphi\nabla^j\nabla_k\varphi\nabla^k\nabla_i\varphi$),
which are commonly encountered in Vainshtein-type theories. In fact,
in the most popular examples of such models -- the DGP, cubic Galileon
and quartic Galileon models -- we also found that the discretisation
could be done in a way such that
$\mathcal{L}^h\left(u_{i,j,k}\right)=0$ is a quadratic or cubic
equation that can be solved analytically. This fact has been used in
\cite{lzk2013, letal2013, blhbp2013, bbl2015} to make simulations of
these models possible, more efficient and free from numerical
instabilities.

Unfortunately, the new method does {\it not} apply to {\it all}
nonlinear partial differential equations, because it relies on
$u_{i,j,k}$ being analytically solvable in the discrete equation. In
the HS $f(R)$ model with $n=1$, $u_{i,j,k}$ satisfies a cubic
equation, which does have analytical solutions. This neat property
does not hold for other models. However, this method will still be
very useful for the following reasons:

\begin{itemize}

\item At the moment, no specific functional forms of $f(R)$ -- or more
  generally, no specific chameleon models -- are known to be
  fundamental. Different models often share similar qualitative
  behaviours though the predictions can be quantitatively
  different. For what it is worth, the HS model serves as a great test
  case to gain insights into the question `How much deviation from GR
  (in the manner prescribed by the large class of chameleon models) is
  allowed by cosmological data?'. Indeed, all current observational
  constraints on modified gravity are to be considered as attempts to
  answer this question. In this context, the exact functional form of
  $f(R)$ is not critical, because whatever form we adopt, it is
  unlikely to be the true theory. Actually, the HS model is capable of
  reproducing the behaviours of many classes of models, and is
  therefore a representative example.

\item {There are other models that this method can be applied to. One
  example is the HS $f(R)$ model with $n=2$. In this case
  Eq.~(\ref{eq:cubic_equation}) becomes a quartic equation, which also
  has analytical solutions. A further example is the logarithmic
  $f(R)$ model studied in the literature \citep[e.g.,
    ref.][]{bbds2008}:

\begin{eqnarray}
f(R) \sim -2\Lambda - \eta\log\left(R/R_\ast\right),\nonumber
\end{eqnarray}

where $\Lambda$ is the cosmological constant, and $\eta$ and $R_\ast$
are some model parameters. In this case, $f_R\sim1/R$, and we could
define $u=-\tilde{f}_R$ so that Eq.~(\ref{eq:cubic_equation}) becomes
a quadratic equation. Moreover, looking beyond $f(R)$ gravity, there
are also other chameleon models with different coupling strengths from
the value of $1/3$ for $f(R)$ models, and can be simulated using this
method \cite{bdlwz2013}. The method can also be applied to the
symmetron model \cite{hk2010}, in which the equation
$\mathcal{L}^h\left(\varphi_{i,j,k}\right)=0$ is a cubic equation
\cite{dlmw2012} for the symmetron field $\varphi$, and certain
variants of the dilaton model \cite{bbds2010,bdlwz2012}, though our
initial tests showed that the improvement of the efficiency is far
smaller than in the $f(R)$ case (Appendix~\ref{app:symmetron}).}

\end{itemize}

Efforts towards generalising the new method to the models mentioned
above, and to running large high resolution simulations including
baryonic physics, are currently ongoing and will be the subject of
future works.

\begin{acknowledgments}
We thank the organisers of the 1st Mexican Numerical Simulations
School held in Mexico City (03-06 October 2016), for inspiring
discussions that motivated this study. We are also grateful to Fabian
Schmidt for many enlightening comments. SB is supported by STFC
through grant ST/K501979/1. BL acknowledges support by STFC
Consolidated Grant Nos. ST/L00075X/1 and RF040365. WAH acknowledges
support from the Polish National Science Center under contract
\#UMO-2012/07/D/ST9/02785. KK is supported by STFC grant ST/N000668/1.
Both WAH \& KK are also supported by the European Research Council
through grant 646702 (CosTesGrav). CLL acknowledges ICRAR (University
of Western Australia) and Queensland University for their warm
hospitality and Chris Power and David Parkinson for discussions.  CLL
also acknowledges support from STFC consolidated grant ST/L00075X/1 as
well as funding support from the University of Queensland - University
of Western Australia Bilateral Research Collaboration Award.  This
work was supported by resources provided by The Pawsey Supercomputing
Centre with funding from the Australian Government and the Government
of Western Australia. The simulations described in this study used the
DiRAC Data Centric system at Durham University, operated by the
Institute for Computational Cosmology on behalf of STFC DiRAC HPC
Facility (www.dirac.ac.uk). This equipment was funded by BIS National
E-Infrastructure Capital grant ST/K00042X/1, STFC Capital grant
ST/H008519/1, and STFC DiRAC Operations grant ST/K003267/1 and Durham
University. DiRAC is part of the National E-Infrastructure. Part of
this work was performed at the Aspen Center for Physics, which is
supported by National Science Foundation grant PHY-1066293.
\end{acknowledgments}

\appendix
\section{The poorer performance of the speed up truncation method in chameleon models}
\label{app:trunc}

In \cite{bbl2015}, the authors proposed a method to speed up $N$-body
simulations of modified gravity models with Vainshtein screening. The
speed up in this method is achieved by truncating the Gauss-Seidel
iterations of the scalar field above a certain refinement level, and
then computing the solution on those fine levels by interpolating from
coarser levels. This {\it approximate} method agrees very well with
the results of the full $N$-body calculation (see \cite{bbl2015} for
details) due to the fact that in Vainshtein screening models, there is
a correlation between higher density regions (or, equivalently, higher
refined regions in the simulation box) and screening efficiency. Even
when the error induced on the fifth force on the refinements is large,
it does not propagate to the total gravitational force because the
amplitude of the fifth force is small/screened. 

In chameleon models, however, the correlation between high density
regions and screening efficiency becomes less marked because of the
dependence on the environmental density (in Vainshtein models, the
screening efficiency depends on the local density only). For example,
in $f(R)$ models, a low mass halo in a dark matter void constitutes an
example of a highly-refined region (the centre of the halo can be very
concentrated) that may not be screened (either by itself or by the low
density environment it lives in). It is therefore interesting to
determine whether or not the same truncation method, which works well
for Vainshtein models, works equally well in chameleon-type theories. 

\begin{figure}
\centering
\includegraphics[width=\columnwidth]{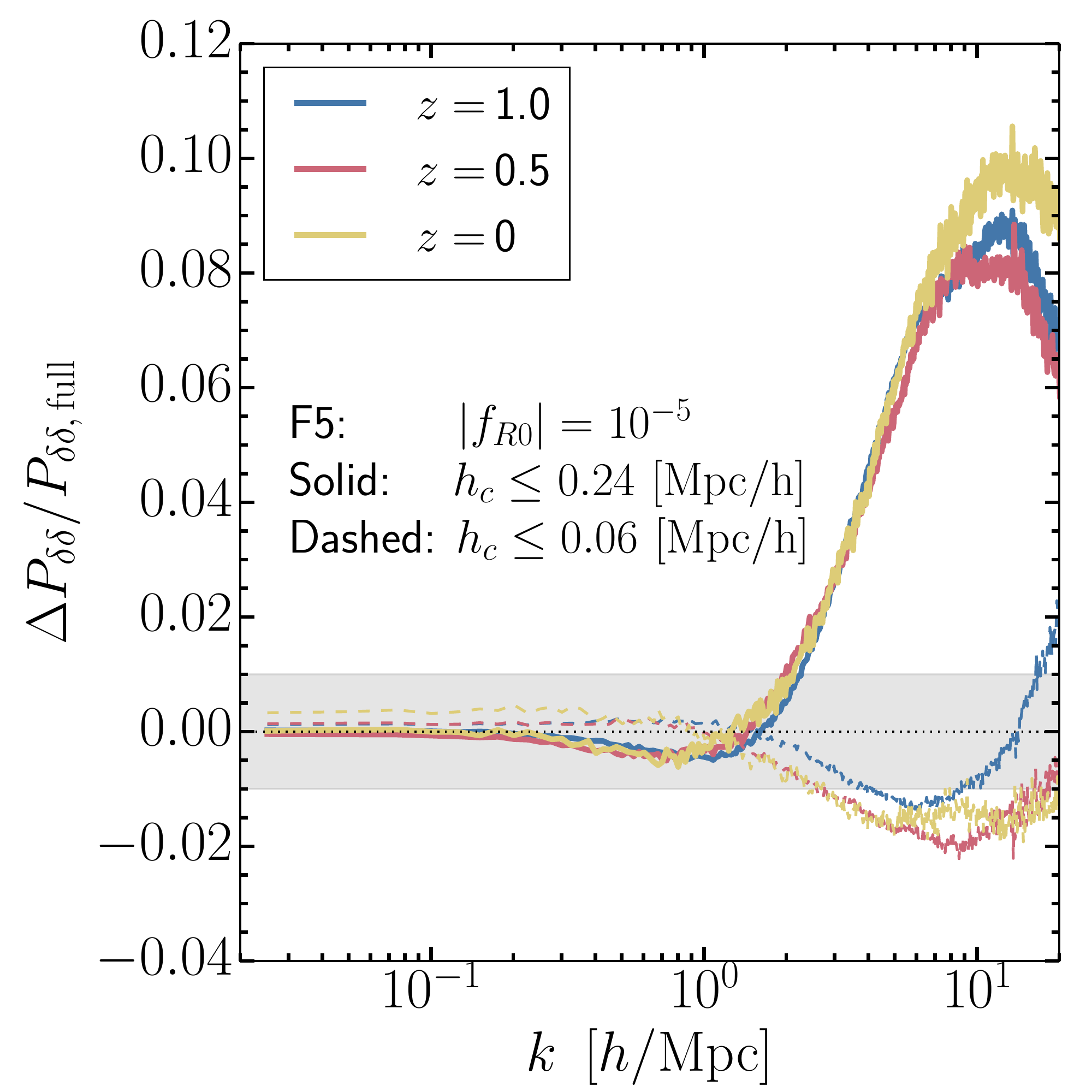}
\caption{Relative difference in the matter power spectra at~
  $z~=~0,~0.5~\&~1$ between a full F5 simulation, and two truncated
  runs where the Gauss-Seidel iterations of the scalar field have been
  truncated on finer refinement levels (see the accompanying
  text). The dashed and solid lines, respectively, correspond to less
  and more aggressive truncation schemes; $h_c$ is the cell size of
  the first truncated level in each simulation. The shaded grey band
  represents the 1\% error region around zero.}
\label{fig:F5_trunc}
\end{figure}

Fig.~\ref{fig:F5_trunc} shows the relative difference of two truncated
$f(R)$ simulations to a full (ie., not truncated) simulation. The
simulation box used for this test is the same as the `Medium res'
setup in the main text, but with $f_{R0} = -10^{-5}$ (the so-called F5
model). The result is shown at three different redshifts and the two
labelled truncation schemes are as follows. The case $h_c \leq
0.24\ {\rm Mpc}/h$ indicates that the scalar field was only explicitly
solved on the coarse level, with this solution being interpolated to
all finer levels. In the case of $h_c \leq 0.06\ {\rm Mpc}/h$, the
scalar field was explicitly solved on the coarse, first refinement and
second refinement levels, with the solution at the second level being
interpolated to all other finer levels. The values $0.24\ {\rm Mpc}/h$
and $0.06\ {\rm Mpc}/h$ indicate the cell size of the first truncated
level in both these simulations, which ran, respectively, $\approx 10$
and $\approx 2$ times faster than the full run. For both these
truncation criteria, the figure shows that the error can be kept $<
1\%$ for $k \lesssim 2\ h/{\rm Mpc}$, but for higher modes, it grows
to unacceptably large values. For example, at $k \approx 5\ h/{\rm
  Mpc}$, the error is of $\approx 6\%$. 

The result shown in Fig.~\ref{fig:F5_trunc} for $f(R)$ should be
contrasted with the corresponding picture in the DGP model (which
employs Vainshtein screening), in which for the same truncation
criteria, the error is always kept below $1\%$ for $k < 5\ h/{\rm
  Mpc}$ (see e.g.,~Fig. 5~of \cite{bbl2015}). Furthermore, the method
described in the main body of this paper results in comparable boosts
in performance compared to previous $f(R)$ simulations, but without
any loss in accuracy. From this we can conclude that the truncation
scheme that works well in Vainshtein screening models is not suitable
for chameleon theories.

\section{Performance of the new method for the symmetron model}
\label{app:symmetron}

As a test of the performance of our new method for other classes of
screening mechanisms, we implemented our method for the case of the
symmetron model. The code used in this case, Isis \citep{isis}, is a
modified version of {\sc ramses} developed independently of {\sc
  ecosmog}. Details of the symmetron model and its implementation in
Isis are described in \cite{isis}. Briefly, the equation of motion for
the scalar field is given by:
\begin{eqnarray}
\nabla^2 \phi \propto \left( A \rho - 1\right) \phi + \phi^3, 
\label{eq:eom_symm}
\end{eqnarray}
where the quantity $A$ is a function of the parameters of the
symmetron model.  While the equation is formally equivalent to the
$f(R)$ in the main text (Eq.~\ref{eq:eom}), the screening mechanism
operates differently. In the $f(R)$ model, the scalar field screens
itself by becoming very massive. On the other hand, in the symmetron
model, the screening occurs when a particular symmetry is restored
(i.e., when the factor in front of the linear term of the source of
Eq.~(\ref{eq:eom_symm}) becomes positive). Consequently, the model
behaves in a different manner to $f(R)$. For instance, negative
solutions for the symmetron field, $\phi$, are allowed and, thus, the
constraints implemented in the $f(R)$ case (\S~\ref{sect:new}) are not
required. We refer the reader to \cite{lp2014} for a summary of the
complex phenomenology associated with this property of the symmetron
field.

The nonlinear modified gravity solver in Isis is very similar to that
of the $f(R)$ model in {\sc ecosmog}. The code uses an implicit
multigrid solver with full approximation storage, which means that the
code relies on a Newton-Raphson algorithm to evolve the solution in
every step of the Gauss-Seidel iterations. As the discretised equation
is cubic, the method proposed in this paper can be applied in a
straightforward manner. As a check of the accuracy of the new method
in solving the symmetron field equations, we have repeated
satisfactorily the static test presented in the original Isis paper
\citep[Fig. 2 in ref.][]{isis}. However, we find that there is no
major difference in the performance of the standard Isis
implementation compared to using the new method, either in terms of
the run time, or the convergence rate of the iterative solver.

In order to gauge the difference in computing time between the old and
new methods for the symmetron model, we have run a set of five
different realisations of a box of size $150 \, {\rm Mpc}/h$ on a
side, containing $256^3$ particles.  For each realisation, there are
three sets of simulations: $\Lambda$CDM and the symmetron model using
the old and new methods. Overall, we do not find any improvement in
the performance of Isis using the new method. For both the old and new
methods, the overhead compared the $\Lambda$CDM simulation is of the
order of $\sim 170\%$ and, in fact, the run time using the new method
is actually $\sim 1\%$ slower than using the default implementation -
this is explained by the fact that $\sim 1\%$ more iterations were
required for the whole set of five realisations using the new
method. The convergence criterion on the residual was set to $10^{-6}$
for both symmetron runs; we find that, unlike in the $f(R)$ model,
making the convergence criterion even stricter does not impact the run
time of the symmetron simulations by a great amount.

The reason why the performance of the code appears to be insensitive
to the details of the iteration scheme is seemingly related to the
type of screening mechanism used by the symmetron model. The symmetron
mechanism is based on a density threshold above which the solution
very quickly approaches zero and thus decouples the scalar field from
matter. This makes the solutions more stable and, therefore, not
strongly dependent on the details of the solver employed. Since the
default solver in Isis does not involve a nonlinear change of
variables to force a stable, positive solution (as in the $f(R)$
case), the performance is already similar to what {\sc ecosmog} can do
for $f(R)$ using the new method.

\end{document}